\documentclass[reprint,superscriptaddress,aps,prl]{revtex4-1}
\usepackage{mathrsfs}
\usepackage{graphicx}
\usepackage{dcolumn}
\usepackage{bm}
\usepackage{amsmath}
\usepackage{amsfonts}
\usepackage{amssymb}
\usepackage{geometry}
\usepackage{makecell}
\usepackage{multirow}%
\usepackage{dcolumn}
\usepackage{epstopdf}
\usepackage[colorlinks,linkcolor=blue,citecolor=blue,urlcolor=blue]{hyperref}
\usepackage[mathlines]{lineno}

\newcommand{\ucite}[1]{\textsuperscript{\cite{#1}}}
\begin{document}


\title{An Origin of Dzyaloshinskii-Moriya Interaction at Graphene-Ferromagnet Interfaces Due to the Intralayer RKKY/BR Interaction}

\author{Jin Yang}
 \affiliation{School of Science and Laboratory of Quantum Information Technology,
Chongqing University of Posts and Telecommunications, Chongqing, 400065, China.}
\author{Jian Li}%
\affiliation{School of Science and Laboratory of Quantum Information Technology,
Chongqing University of Posts and Telecommunications, Chongqing, 400065, China.}
\author{Liangzhong Lin}%
\affiliation{School of Information Engineering, Zhongshan Polytechnic, Zhongshan, 528400, China}

\author{Jia-Ji Zhu}
\email{zhujj@cqupt.edu.cn}
\affiliation{School of Science and Laboratory of Quantum Information Technology,
Chongqing University of Posts and Telecommunications, Chongqing, 400065, China.}%


\begin{abstract}
We present a theory of both the itinerant carrier-mediated RKKY interaction and the virtual excitations-mediated Bloembergen-Rowland (BR) interaction between magnetic moments in graphene induced by proximity effect with a ferromagnetic film. We show that the RKKY/BR interaction consists of the Heisenberg, Ising, and Dzyaloshinskii-Moriya (DM) terms. In the case of the nearest distance, we estimate the DM term from the RKKY/BR interaction is about $0.13$ meV for the graphene/Co interface, which is consistent with the experimental result of DM interaction $0.16\pm0.05$ meV. Our calculations indicate that the intralayer RKKY/BR interaction may be a possible physical origin of the DM interaction in the graphene-ferromagnet interface. This work provides a new perspective to comprehend the DM interaction in graphene/ferromagnet systems.
\end{abstract}

\maketitle


\emph{Introduction.---}Graphene has been the superstar for its unique properties in condensed matter physics since 2004. Magnetism and magnetic phenomena in graphene are the key issues in implementing spintronic devices. Since the intrinsic magnetism is absent in graphene, one may induce magnetism by some extrinsic strategies, for example, creating magnetic moments through introducing vacancies\ucite{125408,n5840,096804}, adding adatoms\ucite{026805,42010,352437}, or doping magnetic impurities\ucite{075414,236602}. However, the feasibility of these approaches is under debate, and the experimental realizations pose tough challenges\ucite{9220,207205,8199}. Therefore, the other extrinsic strategies such as edge engineering in zigzag nanoribbons\ucite{444347,047209,514608}, biased bilayer graphene\ucite{115425,186803}, or magnetic proximity effect borrowed from adjacent magnetic materials\ucite{016603,15711}, receive great attention. The magnetic proximity effect could bring about the strong hybridization between the $2p_z$ orbitals of the carbon atoms with the $d$ states of the metallic substrate and the sublattice-symmetry breaking\ucite{107602}. The proximity effect may also promote the enhancement of the Rashba spin-orbit coupling (RSOC)\ucite{195452} and anomalous Hall effect in graphene-ferromagnetic films\ucite{016603}. Between the magnetic moments borrowed from the ferromagnetic substrate, an indirect magnetic interaction emerges, which is called the Ruderman-Kittel-Kasuya-Yosida (RKKY) interaction\ucite{9699,1645,106893}.

The RKKY interaction is an indirect magnetic interaction mediated by itinerant electrons or holes between magnetic moments, which permits highly controllability thanks to the itinerant carriers. There have been intense researches on the RKKY interaction in graphene using various approaches like the Matsubara Green's function technique\ucite{116802,184430,115119}, the lattice Green's function technique\ucite{165425,035005}, and the exact diagonalization approach\ucite{205416}. Some common conclusions have been reached: i) The RKKY interaction is a usual isotropic Heisenberg-typed interaction in graphene; ii) The range function decays as $1/R^{3}$ with no oscillation in pristine graphene and oscillates with $1/R^{2}$ decay in doped graphene; iii) The RKKY interaction is ferromagnetic between the magnetic moments on the same sublattices and is antiferromagnetic on the opposite sublattices.

\begin{figure}[th]
\begin{center}
    \includegraphics[width=0.5\textwidth]{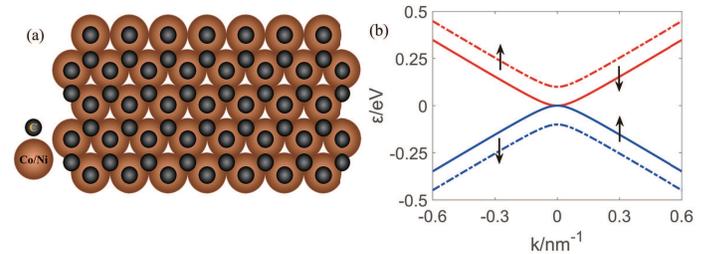}
  \caption{The top-view of the crystal configurations of graphene-coated Ni or Co films in (a), and the (b) is the  band structure for the graphene-nickel system with $\lambda_{R}$=50 meV, $\Delta$=40 $\mu$eV. The red (blue) lines represent conduction (valence) bands, and \rm{$\uparrow$} (\rm{$\downarrow$}) represents spin up (down). }\label{figSchematics}
\end{center}
\end{figure}

 The progress of the spin-orbitronics permits highly efficient electrical control of chiral spin textures---skyrmions or domain wall dynamics, and shows some exciting potential applications in spintronic memory and logic devices\ucite{320190,8152}. These potential applications are based on interfacial Dzyaloshinskii-Moriya (DM) interaction which has been experimentally observed\ucite{67635}. The DM interaction was first proposed for explaining the weak ferromagnetism in oxide materials\ucite{42411958} and is related to spin-orbit coupling\ucite{12091}. Typical DM interaction exists in noncentrosymmetric bulk magnets\ucite{094434,10202} and at interfaces between ferromagnets and metals\ucite{214401,047201,267210,11825}. Yang \textit{et. al.} showed that the physical origin of interfacial DM interaction is the Rashba effect\ucite{17605,014406}. Indeed, several kinds of research shows that there is a large Rashba spin-orbit splitting at the graphene-ferromagnet (\textit{e.g.} Ni or Co) interface\ucite{195452,115426}, and the RSOC could be enhanced to a giant Rashba effect by the intercalation\ucite{057602,31232}. However, the giant Rashba effect could also make some difference to the RKKY interaction in graphene and may result in an anisotropic coupling which consists of the DM interaction\ucite{441538}. Vedmedenko \textit{et. al.} showed that the interfacial DM interaction attributes to an interlayer DM term from the RKKY interaction (L\'{e}vy-Fert model) and the intralayer RKKY interaction is dominant a ferromagnetic Heisenberg term\ucite{257202}. However, we will show that the RKKY interaction also contains an intralayer DM term which contributes to the interfacial DM interaction.

In this paper, we present the RKKY interaction in graphene between magnetic moments induced by proximity effect of a ferromagnetic metal. The ferromagnet provides both broken time-reversal symmetry and RSOC. Due to the RSOC resulting from the graphene-ferromagnet interface, we find that there are Heisenberg, Ising, and DM terms in the RKKY interaction rather than the usual Heisenberg-typed interaction in graphene. This DM interaction could also induce magnetic chirality and weak ferromagnetism. We also consider the Bloembergen-Rowland (BR) interaction when the Fermi energy lies across the neutral point\ucite{971679}, and show that the BR interaction mainly consists of DM term and Heisenberg term in the nearest distance. We estimate the DM interaction arising from the RKKY/BR interaction is about $0.13$ meV for the graphene-Co interface, which is consistent with the experimental result of DM interaction $0.16\pm0.05$ meV\ucite{17605}. Our results indicate that the DM interaction may not be induced by the Rashba effect directly but may be induced by the Rashba effect indirectly via the BR/RKKY interaction. We also show that the RKKY interaction is nearly independent of Fermi energy in the nearest distance, and almost the same as the BR interaction. The Fermi energy hardly influences our estimate of the DM interaction.

\emph{Model.---}We consider two magnetic moments located in graphene, which are borrowed from an adjacent ferromagnetic metal (Co or Ni films). The atomic distance of \textit{fcc} Ni(111) or \textit{hcp} Co(0001) planes is almost perfectly matched with the length of the basis vector of graphene, as showed in Fig~.\ref{figSchematics}(a). The total Hamiltonian can be written as $\mathbf{H=H_{0}}+\mathbf{H}_{\mathrm{int}}$, where $\mathbf{H_{0}}$ is the Hamiltonian of itinerant electrons in graphene with RSOC and intrinsic spin-orbit coupling (ISOC)\ucite{224438,165310},
\begin{align}
\mathbf{H}_{0}&=\hbar v_{F}\left(  \mathbf{\tau}_{x}k_{x}+\mathbf{\tau}_{y}k_{y}\right)  +\lambda
_{R}\left(  \mathbf{\tau}_{x} \mathbf{\sigma}_{y}-\mathbf{\tau}_{y}\mathbf{\sigma}_{x}\right)
  +\Delta \mathbf{\tau}
_{z}\mathbf{\sigma}_{z}\tag{1}%
\label{one}
\end{align}
where the $\lambda_{R}$ and $\Delta$ are the RSOC and the ISOC constant respectively, the $v_{F}$ denotes the Fermi velocity of graphene, Pauli matrices of pseudospin $\mathbf{\tau}$ operate on A(B) sublattices, and $\mathbf{\sigma}$ are Pauli matrices for real electron spin.

The exchange interaction between the local magnetic moments, $\mathbf{S}_{1}$ and
$\mathbf{S}_{2}$, induced by proximity effect of a ferromagnet with the itinerant electron spins $\mathbf{\sigma}$ can be given by\ucite{224438,097201,103008} $\mathbf{H}_{\mathrm{int}}=-J\left(  \mathbf{\sigma}\cdot \mathbf{S}_{i}\right)  \delta \left(\mathbf{r}-\mathbf{R}_{i}\right)$, where $J$ denotes the strength of the $s$-$d$ exchange interaction.

In the loop approximation we find the RKKY interaction between two magnetic moments in the form of\ucite{097201,103008}
\begin{align}
\mathbf{H}_{\alpha \beta}^{RKKY}    &=-\dfrac{J^{2}}{\pi}\operatorname{Im}\int_{-\infty
}^{\varepsilon_{F}}\mathrm{d}\varepsilon \operatorname*{Tr}\left[  \left(  \mathbf{\sigma
}\cdot \mathbf{S}_{1}\right) \right. \nonumber\\
&\times \left.\mathbf{G}_{0}^{\alpha \beta}\left(  \mathbf{R}, %
\varepsilon \right)  \right.
 \left.  \left(  \mathbf{\sigma}\cdot \mathbf{S}_{2}\right)  \mathbf{G}_{0}^{\beta \alpha
}\left(  -\mathbf{R}, \varepsilon \right)  \right] \tag{2}%
\label{two}
\end{align}
where $\alpha,\beta$ are the sublattice indices, A or B, and $\mathbf{R}$ denotes the lattice vector between the sublattice $\beta$ and $\alpha$. Here $\mathbf{G}_{0}^{\alpha \beta}\left(\mathbf{R,}\varepsilon \right)  $ is the unperturbed  Green's function in energy-coordinate representation and $\varepsilon_{F}$ is Fermi energy. Tr means a partial trace over the spin degree of freedom of itinerant Dirac electrons. The Green's functions in real space can be calculated by integrating the corresponding Green's functions in momentum space over the wave vector $\mathbf{k}$ in the vicinity of $K$ valley\ucite{165425,224438}
\begin{align}
\mathbf{G}_{0}^{\alpha \beta}\left(  \pm \mathbf{R,}\varepsilon \right)  =\dfrac{1}%
{\Omega_{\mathrm{BZ}}}\int d^{2}ke^{\pm \mathrm{i}\mathbf{k}\cdot \mathbf{R}}\mathbf{G}_{0}^{\alpha \beta
}\left(  \mathbf{k},\varepsilon \right) \tag{3}%
\label{three}
\end{align}

The $\mathbf{k}$ dependent Green's functions, $\mathbf{G}_{0}\left(  \mathbf{k,}\varepsilon \right)=\left(  \varepsilon+\mathrm{i}\eta-\mathbf{H}_{0}\right)^{-1}$, corresponding to the Hamiltonian (\ref{one}), can be rewritten as a $2\times2$ matrix form in the basis of different sublattices.

With all the calculated on-site Green's functions and site-site Green's functions, the effective RKKY interaction can be obtained as
\begin{align}
\mathbf{H}_{\alpha,\beta}^{RKKY}   &=J_{H}^{\alpha \beta}\mathbf{S}_{1}\cdot
\mathbf{S}_{2}+J_{DM}^{\alpha \beta} \left(  \mathbf{S}%
_{1}\times \mathbf{S}_{2}\right)_{y}\nonumber \\
&+ J_{z}^{\alpha \beta}{S}_{1z}{S}_{2z} +J_{y}^{\alpha \beta}{S}_{1y}{S}_{2y} \tag{4}%
\label{four}
\end{align}

The effective anisotropic spin-spin interaction between local moments includes Heisenberg, Ising, and DM interactions. This behavior of RKKY interaction recurs in topological insulators\ucite{097201}, topological semimetals\ucite{241103,224435}, \textit{p}-doped transition-metal dichalcogenides\ucite{161403} \textit{etc.} The three type terms arise from the symmetry-breaking due to the RSOC and the ISOC.
\begin{table*}[htb]
\caption{The RKKY interaction range function including both of the same sublattices (AA) and the opposite sublattices (AB), where $\kappa{=}%
\frac{2\pi^{2}}{4\hbar^{2}v_{F}^{2}\Omega_{BZ}}, \zeta_{\pm}{=}\sqrt
{\varepsilon^{2}-\Delta^{2}\pm2\left(  \varepsilon-\Delta \right)  \lambda_{R}%
}, \mathscr{J}{=}-\frac{\kappa^{2}J^{2}}{2\pi}$, the $s (s^{\prime})=\mathrm{sgn} (\varepsilon\pm\lambda_{R})$ respectively.}
\label{table2}
\begin{ruledtabular}
\centering
\begin{tabular}{ll}

 \makecell[c]{AA}& \makecell[c]{AB}\\[2pt]
\hline \\ 
 {$J_{y}^{AA}=\mathscr{J} \operatorname{Im}\int_{-\infty
}^{\varepsilon_{F}}\mathrm{d}\varepsilon\left(-4\Gamma_{1}^{2}\right)$}               & {$J_{y}^{AB}=\mathscr{J}\operatorname{Im}\int_{-\infty}^{\varepsilon_{F}%
}\mathrm{d}\varepsilon(-4\Lambda_{2} \Lambda_{3}  )$}                             \\ [5pt]
{$J_{z}^{AA}=\mathscr{J} \operatorname{Im}\int_{-\infty
}^{\varepsilon_{F}}\mathrm{d}\varepsilon\left( -4\Gamma_{3}^{2}\right)$}           &        {$J_{z}^{AB}=\mathscr{J} \operatorname{Im}\int_{-\infty}^{\varepsilon_{F}} \mathrm{d}\varepsilon\left(\Lambda_{2}-\Lambda_{3} \right)^{2}$}                    \\[5pt]
 {$J_{DM}^{AA}=\mathscr{J} \operatorname{Im}\int_{-\infty
}^{\varepsilon_{F}}\mathrm{d}\varepsilon(-4\Gamma_{1}\Gamma_{2})$}                  &
{$J_{DM}^{AB}=\mathscr{J}\operatorname{Im}\int_{-\infty}^{\varepsilon_{F}} \mathrm{d}\varepsilon(-2)\Lambda_{1}\left(\Lambda_{2}+\Lambda_{3} \right)$}         \\[5pt]
$J_{H}^{AA}=\mathscr{J} \operatorname{Im}\int_{-\infty}^{\varepsilon_{F}}\mathrm{d}
\varepsilon 2\left(  \Gamma_{1}^{2}-\Gamma_{2}^{2}+\Gamma_{3}^{2}\right)$           &       {$J_{H}^{AB}=\mathscr{J} \operatorname{Im}\int_{-\infty}^{\varepsilon_{F}}\mathrm{d}\varepsilon2\left(
\Lambda_{2} \Lambda_{3}-\Lambda_{1}^{2} \right)$}                                   \\[5pt]
$\Gamma_{1}=s\zeta_{+}\mathrm{H}_{1}^{\left(  1\right)  }\left(  \tfrac{sR\zeta_{+}}{\hbar
v_{F}}\right)  -s^{\prime}\zeta_{-}\mathrm{H}_{1}^{\left(  1\right)  }\left(  \tfrac{s^{\prime}R\zeta_{-}%
}{\hbar v_{F}}\right)  $                                                              & $\Lambda_{1}=s\zeta_{+}\mathrm{H}_{1}^{\left(  1\right)  }\left(
\tfrac{sR\zeta_{+}}{\hbar v_{F}}\right)  +s^{\prime}\zeta_{-}\mathrm{H}_{1}^{\left(  1\right)
}\left(  \tfrac{s^{\prime}R\zeta_{-}}{\hbar v_{F}}\right)  $                          \\[5pt]
$\Gamma_{2}=\left(  \varepsilon+\lambda_{R}\right)  \mathrm{H}_{0}^{\left(  1\right)
}\left(  \tfrac{sR\zeta_{+}}{\hbar v_{F}}\right)  +\left(  \varepsilon
-\lambda_{R}\right)  \mathrm{H}_{0}^{\left(  1\right)  }\left(  \tfrac{s^{\prime}R\zeta_{-}%
}{\hbar v_{F}}\right)  $                                                                     &      $\Lambda_{2}=-\left(  \varepsilon-\Delta \right)
\mathrm{H}_{0}^{(1)}\left(  \tfrac{sR\zeta_{+}}{\hbar v_{F}}\right)  +\left(
\varepsilon-\Delta \right)  \mathrm{H}_{0}^{(1)}\left(  \tfrac{s^{\prime}R\zeta_{-}}{\hbar v_{F}%
}\right)  $                                                                                     \\[5pt]
$\Gamma_{3}=\left(  \Delta+\lambda_{R}\right)  \mathrm{H}_{0}^{\left(  1\right)
}\left(  \tfrac{sR\zeta_{+}}{\hbar v_{F}}\right)  +\left(  \Delta-\lambda
_{R}\right)  \mathrm{H}_{0}^{\left(  1\right)  }\left(  \tfrac{s^{\prime}R\zeta_{-}}{\hbar
v_{F}}\right)  $                                                                             &                 $\Lambda_{3} =\left(  \varepsilon+\Delta+2\lambda_{R}\right)
\mathrm{H}_{2}^{\left(  1\right)  }\left(  \tfrac{sR\zeta_{+}}{\hbar v_{F}}\right)-\left(  \varepsilon+\Delta-2\lambda_{R}\right)\mathrm{H}_{2}^{\left(  1\right)
}\left(  \tfrac{s^{\prime}R\zeta_{-}}{\hbar v_{F}}\right) $                 \\[5pt]
\end{tabular}
\end{ruledtabular}
\end{table*}

We list the results for two cases---the magnetic moments locate on the same sublattices or the opposite sublattices, and one can see that the $y$ direction Ising term and the DM term only depend on the RSOC. This means that the Rashba effect plays a key role in realizing the DM interaction. The Ising term of the $z$ direction results from both the ISOC and the RSOC\ucite{235206}. All the range functions $J_{i}^{\alpha \beta}$ of the RKKY interaction are listed in Table~\ref{table2}.

\emph{The RKKY/BR interaction.---}The well-preserved linear dispersion of graphene with the RSOC and ISOC is showed in Fig~.\ref{figSchematics}(b), where the splitting energy of subbands is about 2$\lambda_{R}$. In our calculations we used the strength of $s$-$d$ interaction $J$ about $1\ \mathrm{eV}$\ucite{224438}, the Fermi velocity $10^{6}$ m/s\ucite{081406}, and the ISOC of graphene sheet $\Delta=40\ \mathrm{\mu eV}$\ucite{046403}.

\begin{figure}[hbt]
\begin{center}
\includegraphics[width=0.5\textwidth]{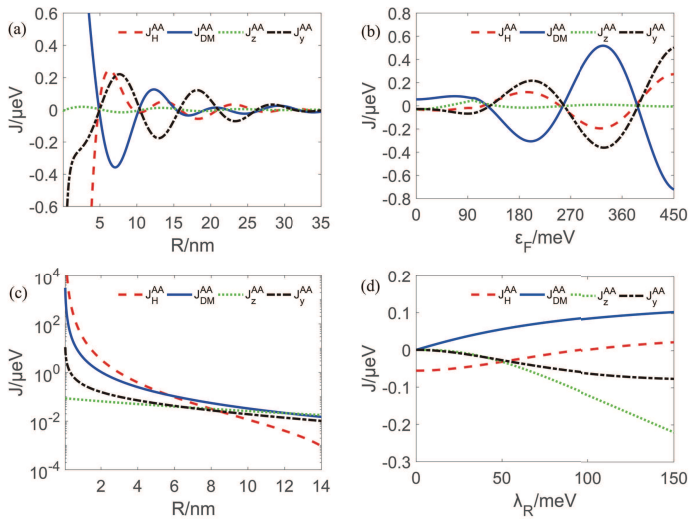}
\caption{All the RKKY interaction range functions of the same sublattice as a function of the distance ${R}$ showed in (a) and Fermi energy $\varepsilon_{F}$ showed in (b). (c) shows the BR interaction of the same sublattices depend on distance in the logarithmic coordinate, and (d) shows the BR interaction of the same sublattices depending on RSOC in units of $\mathrm{meV}$. We set $\Delta=40$ $\mu$eV for all cases, $\lambda_{R}=50 $meV and $\varepsilon_{F}$=200 meV for (a), $\lambda_{R}$=50 $meV$ and $R$=8 nm for (b), $\lambda_{R}=50$ meV and $\varepsilon_{F}$=0 meV for (c), $R$=8 nm, $\varepsilon_{F}$=0 meV for (d).}
\label{fig:2}
\end{center}
\end{figure}

We only need to discuss the case of the same sublattices when we focus on the RKKY interaction in graphene on the Ni(111) or Co(0001) films, because there one carbon atom of the graphene unit cell locates on top of the adjacent Co(Ni) atom and another carbon atom locates above the hollow site, as shown in Fig.~\ref{figSchematics}(a). Generally speaking, the asymptotic behavior of RKKY range functions is determined by the dimensionality of the host materials\ucite{085408}. In 2D electron gas or 2D materials, the range functions decay as $1/R^{2}$ with distance $R$. We can see from Fig. \ref{fig:2}(a) that all the terms decay as $1/R^{2}$ and conform to other 2D systems.

To compare the different terms in the RKKY interaction, we plot all the four types of the RKKY range functions depending on the distance $R$ or Fermi energy $\varepsilon_F$ in Fig. \ref{fig:2}(a) and (b) . We can see the Ising term of the $z$ direction is rather weak and relatively insensible to the increase of the distance between moments, and this is because the Ising term $J_{z}$ relies mainly on the ISOC and the strength of the ISOC is low. The other terms show decaying oscillations with increasing distance, as common RKKY range functions do. Fig.~\ref{fig:2}(b) shows the dependence of the four terms as functions of Fermi energy. Again the Ising term of the $z$ direction shows a relatively flat curve with increasing Fermi energy, and the other range functions show normally enhanced oscillations with the increase of Fermi energy.

While the Fermi energy is cross the neutral point and the density of state drops to zero, there are no carriers in the graphene sheet. The RKKY interaction seems to vanish due to the absence of the mediated carriers. However, the virtual excitations could also offer mediated carriers, and the RKKY interaction would turn to the BR interaction\ucite{971679,097201}. The BR interaction, in general, exponentially decays with increasing the distance between moments. We can see from Fig.~\ref{fig:2}(c) the perfect exponential decaying behavior of the BR interaction. The Heisenberg term falls more rapidly than the other three terms. The intensity of the Heisenberg interaction will be lower than that of the DM interaction in the distance larger than $ 5.7\  \mathrm{nm}$.

In comparison, the intensity of the Ising interaction of the $z$ direction will be higher than that of the DM interaction in the distance larger than $12.1\ \mathrm{nm}$. When the distance approaches zero, the relative strength of two Ising terms almost vanish. However, the Heisenberg term is dominant in this case. In Fig.~\ref{fig:2}(d), we plot the RSOC dependence of the different terms in the BR interaction. We can see the DM term and the Ising terms monotonously increase with increasing RSOC, whereas the isotropic Heisenberg interaction decreases, indicating the anisotropy of the BR interaction is also arise from the RSOC.

\begin{figure}[htb]
\begin{center}
\includegraphics[width=0.5\textwidth]{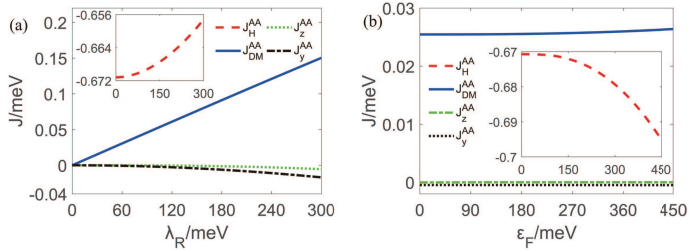}
\caption{(a) The range functions of the BR interaction of the same sublattices depending on RSOC in units of $\mathrm{meV}$ with Fermi energy $ \varepsilon_{F}$=0 meV. (b) The range functions of the RKKY interaction of the same sublattices depending on Fermi energy with RSOC $\lambda_R$=50 meV. $\Delta$=40 $\mu$eV, $R$=0.25 nm  are both for (a) and (b).}
\label{fig:8}%
\end{center}
\end{figure}

\emph{The DM interaction from the intralayer RKKY/BR interaction.---} Since the adjacent Co(Ni) atoms locate under the same sublattice of graphene with a distance of about 0.25 $\mathrm{nm}$\ucite{92844,022509}, we turn our attention to the case of $R=0.25$ nm. We can see from Fig.~\ref{fig:8}(a) that the DM term significantly depends on the RSOC, while the other three terms show insensible dependence. At the nearest distance, the isotropic Heisenberg term makes the major contribution. However, we do not care about the isotropic Heisenberg interaction which offers a usual ferromagnetic exchange interaction. We are interested in the DM interaction arising from the BR or RKKY interaction. Since the parameters of RSOC for Co film $\lambda_R=265$ meV and for Ni film $\lambda_R=136$ meV \ucite{195452}, we can calculate the corresponding DM interaction $J_{DM}=0.133$ meV for graphene/Co and $J_{DM}=0.069$ meV for graphene/Ni from the BR interaction. We compare our estimate of the DM interaction with experimental result $J_{DM}=0.16\pm0.05$ meV of the graphene/Co film\ucite{17605} , and the theoretical estimate is perfectly consistent with the experimental result. This coincidence indicates that the DM interaction may not be induced by the Rashba effect directly but may be induced by the Rashba effect indirectly via the BR interaction.

Fig.~\ref{fig:8}(b) shows the RKKY interaction is nearly independent of Fermi energy. The RKKY interaction is almost the same as the BR interaction even if the Fermi energy drastically increases. The corresponding DM terms from the RKKY interaction ($\varepsilon_F=200$ meV) can be extracted as $J_{DM}=0.134$ meV for graphene/Co and $J_{DM}=0.069$ meV for graphene/Ni, which are almost the same as the DM interaction from the BR interaction. One reason for the inert characteristic lies in the fact that the propagating modes of the mediated carriers reduce within the nearest distance and the mediated carriers involved nearly saturate. The other reason is that the distance is too short to accumulate the angle of spin twisting and the phase changing of the itinerant carriers.

How can we identify the DM interaction from the intralayer RKKY/BR interaction or from the Rashba effect directly? Since the DM interaction from the intralayer BR/RKKY interaction is always along with a Heisenberg term, and we can decide theoretically both the energies of the DM term $J_{DM}$ and the ferromagnetic Heisenberg term $J_{H}$. If we could extract the $J_{DM}$ and $J_{H}$ from asymmetric spin-wave dispersion by measuring the highly resolved spin-polarized electron energy loss spectra\ucite{137203}, we may verify our conclusion by comparing the theoretical value and the experimental result of the $J_{DM}/J_{H}$.

\emph{Conclusion.---}We have studied the long-range RKKY interaction (the short-range BR interaction) mediated by itinerant carriers (virtual excitations) in graphene between magnetic moments induced by proximity effect with a ferromagnetic film. Thanks to the giant RSOC of the graphene-ferromagnet interface, the RKKY/BR interaction consists of the Heisenberg, DM, and Ising terms. Since the adjacent atoms from a ferromagnetic substrate locate under the same sublattice of graphene, we focus on the RKKY/BR interaction on the same sublattice and find that the DM term and the Heisenberg term make the main contribution. While in the case of the nearest distance, we show the RKKY interaction nearly independent of the Fermi energy and is almost the same as the BR interaction. The Heisenberg and the Ising terms are also insensible to the RSOC except for the DM term. We estimate the DM term from the BR/RKKY interaction is about $0.13$ meV for the graphene/Co interface, which is consistent with the experimental result of DM interaction $0.16\pm0.05$ meV. This result indicates that the DM interaction may not be induced by the Rashba effect directly but may be induced by the Rashba effect indirectly via the intralayer BR/RKKY interaction.

This work was supported by the NSFC (Grants No. 11404043, 1160041160), by the key technology innovations project to industries of Chongqing (cstc2016zdcy-ztzx0067) and by Graduate Research Innovation Project of Chongqing (Grants No. CYS18253).

\end{document}